%Paper: hep-ph/9205228
%From: luke@yukawa.ucsd.edu (Michael Luke)
%Date: Thu, 21 May 92 11:16:09 -0700

%
% USES HARVMAC.TEX
%

\input harvmac

%%%%%%%%%%%%%%%%%%%%%%%%%%%%%%%%%%%%%%%%%%%%%%%%%%%%%%%%%%%%%%%%%%%%%%
%
%  UCSD macros to overwrite some of the definitions in harvmac.tex
%  (include after harvmac.tex)
%  last modified 4/92
%
%%%%%%%%%%%%%%%%%%%%%%%%%%%%%%%%%%%%%%%%%%%%%%%%%%%%%%%%%%%%%%%%%%%%%%%
%
% modify the output routine for the little format
%
\ifx\answ\bigans
\else
\output={
  \almostshipout{\leftline{\vbox{\pagebody\makefootline}}}\advancepageno
}
\fi
%
%
% address
%
\def\mayer{\vbox{\sl\centerline{Department of Physics 0319}%
\centerline{University of California, San Diego}
\centerline{9500 Gilman Drive}
\centerline{La Jolla, CA 92093-0319}}}
%
% grant numbers
%
\def\doe{\#DOE-FG03-90ER40546}
\def\pyiam{PHY-8958081}

%
% preprint number
%
\def\UCSD#1#2{\noindent#1\hfill #2%
\bigskip\supereject\global\hsize=\hsbody%
\footline={\hss\tenrm\folio\hss}}% restores pagenumbers
%
% abstract
%
\def\abstract#1{\centerline{\bf Abstract}\nobreak\medskip\nobreak\par #1}
%
%
% titlefont
%
%
\edef\tfontsize{ scaled\magstep3}
 \tfontsize  \tfontsize
 \tfontsize \font\titlei=cmmi10 \tfontsize
\font\titleis=cmmi7 \tfontsize \font\titleiss=cmmi5 \tfontsize
\font\titlesy=cmsy10 \tfontsize \font\titlesys=cmsy7 \tfontsize
\font\titlesyss=cmsy5 \tfontsize  \tfontsize
\skewchar\titlei='177 \skewchar\titleis='177 \skewchar\titleiss='177
\skewchar\titlesy='60 \skewchar\titlesys='60 \skewchar\titlesyss='60
%
%\def\titlefont{\def\rm{\fam0\titlerm}% switch to title font
%\textfont0=\titlerm \scriptfont0=\titlerms \scriptscriptfont0=\titlermss
%\textfont1=\titlei \scriptfont1=\titleis \scriptscriptfont1=\titleiss
%\textfont2=\titlesy \scriptfont2=\titlesys \scriptscriptfont2=\titlesyss
%\textfont\itfam=\titleit \def\it{\fam\itfam\titleit}\rm}
%
%
% math symbols
%
%---------------------------------------------------------------------
%
\def\inv{^{\raise.15ex\hbox{${\scriptscriptstyle -}$}\kern-.05em 1}}
  %prime
\def\lbar{{\lower.35ex\hbox{$\mathchar'26$}\mkern-10mu\lambda}} %lambda bar

%
%
% various slashed symbols
%
%
\def\slash#1{\rlap{$#1$}/} % slashes a character
\def\dsl{\,\raise.15ex\hbox{/}\mkern-13.5mu D} %this one can be subscripted
\def\delsl{\raise.15ex\hbox{/}\kern-.57em\partial}
\def\Ksl{\hbox{/\kern-.6000em\rm K}}
\def\Asl{\hbox{/\kern-.6500em \rm A}}
\def\Dsl{\hbox{/\kern-.6000em\rm D}} %roman D
\def\Qsl{\hbox{/\kern-.6000em\rm Q}}
\def\gradsl{\hbox{/\kern-.6500em$\nabla$}}
%
% space and backspace in l mode
%
\def\lspace{\ifx\answ\bigans{}\else\qquad\fi}
\def\lbspace{\ifx\answ\bigans{}\else\hskip-.2in\fi} % $$\lbspace...$$
%
%     boxes an equation
%
\def\boxeqn#1{\vcenter{\vbox{\hrule\hbox{\vrule\kern3pt\vbox{\kern3pt
        \hbox{${\displaystyle #1}$}\kern3pt}\kern3pt\vrule}\hrule}}}
%
%     draw a little box (end of proof symbol)
%     e.g. \mbox{.1}{.1}
%
\def\mbox#1#2{\vcenter{\hrule \hbox{\vrule height#2in
\kern#1in \vrule} \hrule}}
%
%
%
%     curly letters
%
   %curly letters
\def\CA{{\cal A}}   
   
   \def\CL{{\cal L}}
  \def\CO{{\cal O}} 
   
\def\CU{{\cal U}} \def\CV{{\cal V}}  
 
%
%
%
%     derivatives
%
%

%

\def\bar#1{\overline{#1}}

\def\bra#1{\left\langle #1\right|}
\def\ket#1{\left| #1\right\rangle}
\def\abs#1{\left| #1\right|}

\def\darr#1{\raise1.5ex\hbox{$\leftrightarrow$}\mkern-16.5mu #1}

%
 %pound sterling
%
 %puts a small half in a displayed eqn
\def\frac#1#2{{\textstyle{#1\over #2}}} %puts a small fraction
%in a displayed eqn
%
%
%     various math operators
%
%

%
%
%
%

%
%       relations
%
\def\ltap{\ \raise.3ex\hbox{$<$\kern-.75em\lower1ex\hbox{$\sim$}}\ }
\def\gtap{\ \raise.3ex\hbox{$>$\kern-.75em\lower1ex\hbox{$\sim$}}\ }
\def\gl{\ \raise.5ex\hbox{$>$}\kern-.8em\lower.5ex\hbox{$<$}\ }
\def\roughly#1{\raise.3ex\hbox{$#1$\kern-.75em\lower1ex\hbox{$\sim$}}}
%
%
%       This defines et al., i.e., e.g., cf., etc.
\def\ie{\hbox{\it i.e.}}

\def\np#1#2#3{{Nucl. Phys. } B{#1} (#2) #3}
\def\pl#1#2#3{{Phys. Lett. } {#1}B (#2) #3}

\relax

\def\leff{\CL_{\rm eff}}

\def\mlimit{m\rightarrow\infty}
\def\vq{v - {q\over m}}
\def\vqp{v + {q\over m}}
\def\av#1{A^{#1}_v}
\def\aw#1{A^{#1}_w}
\def\cav#1{\CA^{#1}_v}

\def\vmeasure{\sum_v\ }
\def\wt{\widetilde}
\def\({\left(}
\def\){\right)}
\def\[{\left[}
\def\]{\right]}
\def\hc{{c_v}}
\def\hb{{b_{v'}}}
\def\hbv{{\hb}}
\def\hcvbar{{\overline{\hc}}}
\def\Dslash{D\hskip-0.6em /}

\def\back#1{\raise1.5ex\hbox{$\leftarrow$}
\mkern-16.5mu #1}

\centerline{{\titlefont{Reparameterisation Invariance Constraints}}}
\medskip
\centerline{{\titlefont{on Heavy Particle Effective Field Theories}}}
\bigskip
\centerline{Michael Luke and Aneesh V. Manohar}
\bigskip
\mayer
\vfill
\abstract{
Since fields in the heavy quark effective theory are described by
both a velocity and a residual momentum, there is redundancy
in the theory:  small shifts in velocity may be absorbed into
a redefinition of the residual momentum.  We demonstrate that this
trivial reparameterisation invariance has non-trivial consequences:
it relates coefficients of terms of different
orders in the $1/m$ expansion and requires linear combinations of these
operators to be multiplicatively renormalised.
For example, the operator $-D^2/2m$ in the effective lagrangian has
zero anomalous dimension, coefficient one, and does not receive any
non-perturbative contributions from matching conditions.
We also demonstrate
that this invariance severely restricts the forms of operators which
may appear in chiral lagrangians for heavy particles.
}
\vfill
\UCSD{UCSD/PTH 92-15}{May 1992}

\newsec{Introduction}
The dynamics of heavy particles at low energies may be described by a heavy
particle effective field theory, in which the effective lagrangian is expanded
in inverse powers of the heavy particle mass
\ref\iw{N.~Isgur and M.~B.~Wise, \pl{232}{1989}{113};
\pl{237}{1990}{527}.}--%OD
\nref\pw{H.~D.~Politzer and M.~B.~Wise, \pl{206}{1988}{681};
\pl{208}{1988}{504}.}%
\nref\vs{M.~B.~Voloshin and M.~A.~Shifman, Yad Fiz. {45} (1987)
463 [Sov.
J. Nucl. Phys. {45} (1987) 292]; Sov. J. Nucl. Phys. {47}
(1988) 511.}%
\nref\eh{E.~Eichten and B.~Hill, \pl{234}{1990}{511}.}%
\nref\grin{B.~Grinstein, \np{339}{1990}{253}.}%
\nref\georgi{H. Georgi, \pl{240}{1990}{447}.}%
\ref\fggw{A.~F.~Falk, H.~Georgi, B.~Grinstein and M.~B.~Wise,
\np{343}{1990}{1}.}. The particles in the effective theory are described by
velocity dependent fields \georgi\ with velocity $v$, residual momentum $k$,
and total momentum $p=mv+k$.
There is an ambiguity in assigning a velocity and momentum to a
particle when one considers $1/m$ corrections to the effective field theory.
The same physical momentum may be parameterised by
\eqn\ireparam{
\left(v,k\right) \leftrightarrow \left(\vqp, k-q\right),\qquad
v^2=\left(\vqp\right)^2 =1,
}
where $q$ is an arbitrary four-vector which satisfies $(v+q/m)^2=1$. The
effective field theory must be invariant under the reparameterisation of the
velocity and momentum, Eq.~\ireparam.  This invariance has long been
recognised (see, for example, \ref\dgg{M.~J.~Dugan, M.~Golden and
B.~Grinstein, HUTP-91/A045, BUHEP-91-18, SSCL-Preprint-12 (1991).}),  however
what is less well known is that it places constraints on the effective
lagrangian, and relates coefficients of terms which are of different order in
the $1/m$ expansion.

We will first discuss the consequences of reparameterisation invariance for the
simple case of a spin-0 field in Sec.~2, and generalise the result to the
somewhat more complicated case of particles with spin in Sec.~3. A
few sample applications to matching conditions, anomalous dimensions and chiral
lagrangians are discussed in Sec.~4.
One important result that is obtained in
Sec.~4
is that the coefficients of certain $1/m$
operators in the effective theory are exactly fixed, and cannot be
modified by non-perturbative corrections.  Section~5 discusses the
consequences of reparameterisation invariance for matrix elements.

\newsec{Reparameterisation Invariance for Scalar Fields}
Consider a coloured
scalar field \ref\gw{H.~Georgi and M.~B.~Wise, \pl{243}{1990}{279}.}
with mass $m$ coupled to gluons, with lagrangian
\eqn\islag{
\CL = D_\mu\phi^* D^\mu \phi - m^2 \phi^* \phi.
}
The low energy effective lagrangian is given in terms of a velocity
dependent effective field \georgi
\eqn\ivfield{
\phi_v(x) = \sqrt{2m}\ e^{i m v\cdot x} \phi(x),
}
where $v$ is a velocity four-vector of unit length, $v^2=1$.
The field $\phi_v$ creates and annihilates scalars with definite velocity $v$,
which is a good quantum number in the $\mlimit$ limit.
The effective lagrangian which describes the low-energy dynamics of the full
theory Eq.~\islag\ is
\eqn\ielag{
\leff = \vmeasure\  \phi^*_v
\left(i v\cdot D\right) \phi_v + \CO\({1\over m}\).
}
The reparameterisation transformation corresponding to
Eq.~\ireparam\ for  the velocity dependent fields is
\eqn\ifredef{
\phi_w(x) = e^{i q \cdot x} \phi_v(x),\qquad w=\vqp,
}
under which the effective lagrangian must remain invariant. We will
explicitly work out the consequences of reparameterisation invariance up
to order $1/m$.
The most general effective lagrangian for the scalar field theory up to
terms of
order $1/m$ is
\eqn\igenlag{
\leff = \vmeasure\
\phi^*_v \left(i v\cdot D\right) \phi_v
 -   {A\over 2 m}\ \phi^*_v\, D^2 \phi_v,
}
where $A$ is a constant, and we have used the lowest order equation
of motion to eliminate a term of the form $\phi^*_v\(v\cdot D\)^2 \phi_v$.
Substituting the reparameterisation transformation \ifredef\ gives
\eqn\inewlag{
\leff = \vmeasure\
\phi^*_w  \left\{
v\cdot (iD + q)\right\} \phi_w
- {A\over 2 m}\ \phi^*_w\ (D^\mu -i q^\mu)^2 \phi_w.
}
Relabelling
the dummy variable $w$ in Eq.~\inewlag\ as $v$ and $v$ as
$v-q/m$, gives the modified lagrangian
\eqn\imodlag{
\leff = \vmeasure
\phi^*_v  \left\{(\vq) \cdot (iD + q)\right\}
 \phi_v
- {A\over 2 m}\ \phi^*_v\ (D -i q)^2\phi_v.
}
Expanding to first order in the (infinitesimal) transformation parameter $q$
gives the change in $\CL$,
\eqn\ichnglag{\eqalign{
\delta\leff &=
\vmeasure
\phi^*_v  \left\{v \cdot q - {i q \cdot D\over m}\right\}
 \phi_v +i{A\over m}\ \phi^*_v\ \(q\cdot D\)\phi_v\cr
&=\left(A-1\right)\phi^*_v\ {q\cdot D\over m}\phi_v
+\CO\left(q^2,{1\over m^2}\right),\cr
}}
using $q\cdot v = \CO(q^2/m)$ from \ireparam. The
lagrangian \igenlag\ is reparameterisation
invariant up to order $1/m$
only if $A=1$. Thus reparameterisation
invariance has
fixed the coefficient of one of the $1/m$ terms in the effective lagrangian.
The tree-level matching condition of Eq.~\ielag\ determined $A=1$, but we now
have the stronger result that $A=1$ is  exact.

It is an elementary exercise to determine the most general possible
reparameterisation invariant scalar lagrangian. The most general possible
lagrangian may be written in the form
\eqn\imglag{
\CL = \vmeasure \CL_v(\phi_v(x),v^\mu,iD^\mu),
}
where $D^\mu$ represents a covariant derivative acting on the heavy field
$\phi_v$.
Substituting the field reparameterisation Eq.~\ifredef, and replacing
the dummy index $w$ by $v$ as before gives
\eqn\imglag{
\CL = \vmeasure \CL_v(\phi_v(x),\(\vq\)^\mu,iD^\mu+q^\mu).
}
For the lagrangian to be reparameterisation invariant, it is necessary and
sufficient that factors of $v$ and $D$ occur only in the combination
\eqn\iinvcomb{
\CV_\mu = v_\mu + {i D_\mu\over m}.
}
This linear combination is precisely $p^\mu/m$, where $p^\mu$ is the total
momentum of the particle, and is the only quantity which is unambiguously
defined at order $1/m$.

The results just derived may be easily extended to include scalar fields
coupled to an external source. The source is velocity independent, and in the
effective theory, it must couple only to reparameterisation invariant
combinations of operators in the effective theory. Thus a scalar source
coupling $J^*(x)\phi(x) $ can couple to
$J^*(x)e^{-i m v \cdot x} \phi_v(x)$ as well as higher dimension operators,  a
vector source can couple to $e^{-i m v \cdot x}
\CV_\mu\phi_v(x)$, and so forth.

\newsec{Vector and Spinor Fields}

The preceding analysis also applies to particles with spin.  The only
complication which arises is that the effective fields satisfy
the velocity dependent constraints
\eqn\consone{{1-\slash v \over 2}\psi_v=0}
for a heavy spinor $\psi_v$, and
\eqn\constwo{{v_\mu\av{\mu}=0}}
for a heavy vector field $\av{\mu}$\ref\chris{C.~Carone,
\pl{253}{1991}{408}.}, which must be preserved by the reparameterisation
transformation.  This makes the transformation law for the fields
somewhat more complicated.

We first consider the case of a heavy vector field $\av{}$.  The
lagrangian must be invariant under the transformation
\eqn\iivrep{
\aw\mu(x) = e^{i q \cdot x} R^\mu{}_\nu(w,v)\av\nu(x),
\qquad w = \vqp,
}
where $R^\mu{}_\nu(w,v)$ is a Lorentz transformation whose form we
must determine.
Define the matrix $\Lambda(v',v)$ to be a Lorentz transformation in the
$v-w$ plane
which rotates $v$ into $v'$, \ie\ $v'=\Lambda(w,v)\, v$.
The $\Lambda$ matrix may be written as
\eqn\iilform{
\Lambda(w,v) = \exp\left[i J_{\alpha\beta} v'^\alpha v^\beta \theta\right],
}
where $\theta$ is the boost angle, and
\eqn\iirmatrix{
\left[J_{\alpha\beta}\right]_{\mu\nu}=-i\left(g_{\alpha\mu}
g_{\beta\nu}-g_{\alpha\nu} g_{\beta\mu}\right)
}
are the Lorentz generators in the spin-1 representation. The Lorentz boost
matrix is computed in Appendix~A.
Consider an external state in the full theory with polarisation vector
$\epsilon$, satisfying $p\cdot\epsilon=0$.  In the effective theory,
the polarisation vectors are given by
\eqn\iieffpol{
\epsilon_v = \Lambda(v,p/m)\, \epsilon,\qquad \epsilon_w = \Lambda(w,p/m)
\, \epsilon,}
so the appropriate reparameterisation transformation for spin-1 fields is
\eqn\iiiapprop{\epsilon_w=\Lambda(p/m,w)\inv\Lambda(p/m,v)
\epsilon_v.}
Note that because the Lorentz group is non-Abelian, this is {\sl not}
the same as the (incorrect) transformation
\eqn\iiincorrect{
\epsilon_w = \Lambda(w,v) \epsilon_v.}
Eqs.~\iiiapprop\ and \iiincorrect\ differ by a Thomas precession term
proportional to $q^{[\alpha}k^{\beta]}/m^2$, the area of the spherical
triangle on $S^3$ with vertices at $v$, $w$ and $p/m$.
It is not possible to make a reparameterisation invariant lagrangian
using the transformation law of Eq.~\iiincorrect. The lagrangian may be
made invariant at order $1/m$ using \iiincorrect, but at order $1/m^2$, there
are terms which are antisymmetric in $D^\mu q^\nu$, which cannot be cancelled
by the variation of any term of order $1/m^2$ in $\leff$.

The transformation \iiiapprop\ is defined for polarisation vectors.  To
find the corresponding field redefinition, $p/m$  should be replaced by
the operator
\eqn\preplace{
p^\mu /m\rightarrow  \CU^\mu,\qquad \CU^\mu = \CV^\mu/\abs{\CV}
}
in Eqs.~\iieffpol\ and \iiiapprop, where $\CV$ is defined in
Eq.~\iinvcomb. The
reparameterisation transformation Eq.~\iiiapprop\ can be written as
\eqn\xxxrep{\eqalign{
\aw{}(x)&=e^{iq\cdot x}\,\Lambda\left(
{ v^\mu + {i D^\mu\over m}\over \abs{v^\mu + {i D^\mu\over
m}}},w\right)\inv\Lambda\left({ v^\mu + {i D^\mu\over m}\over \abs{v^\mu + {i
D^\mu\over m}}},v\right)
\av{}(x)\cr
&=\Lambda\left(
{ w^\mu + {i D^\mu\over m}\over \abs{w^\mu + {i D^\mu\over m}}},w\right)\inv
e^{iq\cdot x}\,\Lambda\left({ v^\mu + {i D^\mu\over m}\over \abs{v^\mu + {i
D^\mu\over m}}},v\right)
\av{}(x),\cr
}}
since
\eqn\xxxcov{
\left(w^\mu + {i D^\mu\over m}\right) e^{iq\cdot x}
= e^{iq\cdot x} \left(v^\mu + {i D^\mu\over m}\right).
}
Thus the  only operator transformation that is required is of the form
\eqn\xxxop{
\Lambda\left({ v^\mu + {i D^\mu\over m}\over \abs{v^\mu + {i D^\mu\over
m}}},v\right)
}
where the same velocity $v$ occurs in both arguments. There is an operator
ordering ambiguity in the transformation Eq.~\xxxop\ at order $1/m^2$,
since
\eqn\xxxnoncom{
\left[\CV^\mu,\CV^\nu\right] = i g {F^{\mu\nu}\over m},
}
which produces an ordering ambiguity in the reparameterisation
transformation Eq.~\iiiapprop\ at order $1/m^3$.
However, different orderings
just differ by powers of the field strength $F^{\mu\nu}$ times $\av{}$,
and correspond to field redefinitions in the effective theory. Thus one can
pick a particular ordering in the definition of $\Lambda$ in Eq.~\xxxop\ and
use it consistently. To order $1/m$, the field $\cav{}$ that appears in the
effective lagrangian is
\eqn\xxxca{
\cav\mu = \av\mu - v^\mu {i D\cdot \av{}\over m}+ \CO\({1\over m^2}\),
}
using Eq.~(A.6) and $v\cdot \av{}=0$.

To construct the most general lagrangian invariant under \iivrep,
it is convenient to introduce the field
\eqn\iiica{
\CA_v^\mu(x) = \Lambda^\mu{}_\nu(p/m,v)\av\nu(x)}
which simply picks up a phase under reparameterisation
\eqn\iiisimple{
\CA_w^\mu(x) = e^{i q\cdot x}\, \CA_v^\mu(x)
}
and satisfies
\eqn\iiisatisfies{
p^\mu\CA_\mu(x)=0.
}
The most general reparameterisation invariant lagrangian may now be
written in the form
\eqn\iiilag{
\CL = \vmeasure\ \CL_v(\CA_v(x),\CV^\mu)=\vmeasure
\CL_v(\Lambda^\mu{}_\nu(p/m,v)\av\nu(x),\CV^\mu),
}
using the same argument as for scalar fields.

Heavy fermions in the
effective theory are described by velocity dependent spinor fields
$\psi_v$ that satisfy the constraint
\eqn\ivconst{
\slash v\ \psi_v = \psi_v
}
(we treat here only the case of fermions; the arguments are easily
generalised to heavy anti-fermions, which satisfy $\slash v\ \psi_v
= - \psi_v$).
A consistent reparameterisation transformation for spinor fields is
defined by analogy with the vector transformation, Eq.~\iiiapprop,
\eqn\ivsptrans{
\psi_w(x) = e^{iq\cdot x}\,
\wt \Lambda(w,p/m)
\wt\Lambda(v,p/m)^{-1}\psi_v(x),\quad w=\vqp,
} where $\wt\Lambda$ are the Lorentz boosts in the spinor
representation. The spinor lagrangian may be written in the form
\eqn\ivsplag{
\CL = \vmeasure\ \CL_v(\Psi_v(x),\CV^\mu),
}
where the reparameterisation covariant spinor field
\eqn\ivcov{
\Psi_v(x) \equiv \wt\Lambda(p/m,v)\psi_v(x),
} transforms as
\eqn\ivcovtrans{
\Psi_w(x) = e^{iq\cdot x}\, \Psi_v(x).
}
The field $\Psi$ may be written using the explicit form for
$\wt\Lambda$ in Appendix~A, and choosing a particular operator ordering
for the covariant derivatives. At order $1/m$,
\eqn\ivPsi{
\Psi_v(x) = \left( 1 +
{i\dsl\over 2m}\right) \psi_v(x).
}
The terms in the effective lagrangian are bilinears in the Fermi fields. The
reparameterisation invariant combinations of the standard fermion
bilinears are
\eqn\ivinvcomb{\eqalign{
\bar\Psi_v \Psi_v & = \bar\psi_v\psi_v,\cr
\bar\Psi_v \gamma_5 \Psi_v & = 0,\cr
\bar\Psi_v \gamma^\mu \Psi_v & = \bar\psi_v \left(v^\mu + {i D^\mu\over
m}\right)\psi_v +\CO\(1/m^2\),\cr
\bar\Psi_v \gamma^\mu\gamma_5 \Psi_v & = \bar\psi_v \(\gamma^\mu\gamma_5 -
v^\mu {i\dsl\over m}\gamma_5\)
\psi_v + \CO\(1/m^2\),\cr
\bar\Psi_v \sigma^{\alpha\beta} \Psi_v & =
\epsilon^{\alpha\beta\lambda\sigma}
 \bar\Psi_v \gamma_\sigma\gamma_5
\CV_\lambda\Psi_v.\cr
}}

\newsec{Applications}

Reparameterisation invariance constrains terms in the effective
lagrangian.  As a simple example, we have already seen that the kinetic
term in the effective theory must have the form
\eqn\vkin{
v \cdot iD + {\left(iD\right)^2\over 2m}, } a result which was proved in
Sec.~2 for scalar fields, but can also be seen to be true for vector
and spinor
fields using the results of Secs.~3--4.  The coefficient of the the
$\left(iD\right)^2$ operator in the effective theory is fixed to be
$1/2m$, and is not renormalised. This agrees with a one loop computation
of the anomalous dimension
\ref\ehgfl{E.~Eichten and B.~Hill, \pl{243}{1990}{427}\semi
A.~F.~Falk, B.~Grinstein and M.~E.~Luke, \np{357}{1991}{185}.}.
More importantly, this result is a non-perturbative
non-renormalisation
theorem. It has recently been suggested that there may be
non-perturbative corrections in the heavy quark theory
\ref\rome{L.~Maiani, G.~Martinelli and C.~T.~Sachrajda, SHEP 90/91-32,
ROME Prep. 812 (1991).} at
order $1/m$ that modify the matching condition for the operator $D^2/m$.
This cannot be true if the effective theory is regulated to preserve
reparameterisation invariance.\footnote{$^\star$}{We thank Mark Wise for
discussions on this point.}.

As another example, the leading spin dependent term in the heavy quark
effective theory is
\eqn\vspin{
{g C\over 2m}\ \bar\psi_v\ \sigma^{\alpha\beta}F_{\alpha\beta}\ \psi_v =
{g C\over 2m}\  \epsilon^{\alpha\beta\lambda\sigma} \bar\psi_v\ v_\lambda
\gamma_\sigma\gamma_5 F_{\alpha\beta}\ \psi_v,
}
where $C=1$ at tree level.  This operator is not related to the kinetic
term by reparameterisation invariance, so $C$ is not protected from
radiative corrections.
Using the results of Eq.~\ivinvcomb, one
finds that the reparameterisation invariant generalisation of Eq.~\vspin
\ to order $1/m^2$ is
\eqn\vnext{\eqalign{
&{gC\over 2m}\ \epsilon^{\alpha\beta\lambda\sigma}
\bar\psi_v\ F_{\alpha\beta} \left(v_\lambda+ {iD_\lambda\over m}
\right) \gamma_\sigma \gamma_5\ \psi_v\cr
=&{gC\over 2m}\ \bar\psi_v\ \sigma^{\alpha\beta}\left(F_{\alpha\beta}
+ 2
F_{\sigma\alpha} {iD_\beta\over m}v^\sigma
\right)\ \psi_v.
}}

A similar analysis applies to external currents in the effective theory.
For example, the weak current $J^\mu=\hcvbar\Gamma^\mu\hbv$, where
$\Gamma^\mu=\gamma^\mu$ or $\gamma^\mu\gamma_5$, and $\hc$ and $\hb$ are
heavy $c$ and $b$ quark fields is written in
reparameterisation-invariant form as
\eqn\vicurr{\eqalign{J^\mu
&=\hcvbar\Gamma^\mu\hbv-{1\over 2 m_c}\hcvbar i\back{\Dslash}\Gamma^\mu
\hbv
%&\quad -{1\over 8 m_c^2}\hcvbar\back{D}^2\Gamma^\mu\hbv
+\CO\({1\over
m_c^2}\)\cr
&+{1\over 2 m_b}\hcvbar\Gamma^\mu\Dslash \hbv
%-{1\over 8 m_b^2}\hcvbar\Gamma^\mu D^2\hbv
+\CO\({1\over
m_b^2}\),}}
This agrees with the results in \ref\chogrin{P.~Cho and B.~Grinstein,
SSCL-Preprint-111, HUTP-92/A012 (1992).}, in which the $\CO(\alpha_s)$ matching
of the operators $\hcvbar\Gamma^\mu\hbv$ and $(-i/2m_c)\hcvbar
\back{\Dslash}\Gamma^\mu
\hbv$ were found to be identical.
It also agrees with \ref\ml{M.~Luke, \pl{252}{1990}{447}.}\ where it was found
that the operators
$\hcvbar\Gamma^\mu\hbv$ and $\hcvbar\back{D}\Gamma^\mu\hbv$ have
the same anomalous dimension in the effective theory.  Furthermore,
it extends this result to additional operators at all orders in
$1/m$.  Note that this does not mean that \vicurr\ is the complete
expression for the current in the effective theory.  There will be other
terms whose coefficients are unrelated to the zeroth order coefficient
by reparameterisation invariance, just as the quark magnetic moment
operator is not determined from the zeroth order kinetic term in the
effective lagrangian.

Finally, reparameterisation invariance also provides useful information
for chiral perturbation theory for heavy
matter fields \ref\jm{E.~Jenkins and A.~V.~Manohar, \pl{255}{1991}{558},
Baryon Chiral Perturbation Theory, UCSD preprint UCSD/PTH 91-30
(1991).}--%
\nref\wise{M.~B.~Wise, Caltech preprint CALT-68-1765 (1992).}%
\nref\bd{G.~Burdman and J.~Donoghue, U.~Mass.~preprint UMHEP-365
(1992).}%
\nref\ycclly{T.-M.~Yan, H.-Y.~Cheng, C.-Y.~Cheung,
G.-L.~Lin, Y.~C.~Lin and H.-L.~Yu, Cornell preprint CLNS 92/1138
(1992).}%
\nref\peter{P.~Cho, HUTP-92-A014 (1992).}%
\ref\fl{A.~F.~Falk and M.~Luke, UCSD/PTH 92-14, SLAC-PUB-5812
(1992).}.
In this case, one cannot
compute the matching conditions explicitly, so the operator coefficients
are undetermined constants. Reparameterisation invariance eliminates a
large number of operators in the chiral expansion, or determines their
coefficients, thus considerably reducing the number of free parameters in
the computation. As a simple example, consider a theory with a heavy
scalar $T_v$ and a heavy vector $B_v^\mu$. The effective lagrangian
could contain a term of the form
\eqn\vieg{
T_v\, iD_\mu\, B_v^\mu.
}
Under the reparameterisation transformation, this term has a variation of
the form
\eqn\viegvar{
T_v\, q_\mu\, B_v^\mu
}
which cannot be cancelled by any term in the effective lagrangian which
is of order one (or of higher order in $1/m$).
This is easily seen by writing the lagrangian
in terms of the fields in \iiilag, where \vieg\ could only arise from
$$
T_v\, \CV_\mu\, {\cal B}_v^\mu
$$
which is zero by \iiisatisfies. Thus the term $T_v\, iD_\mu\, B_v^\mu$
cannot occur in the chiral lagrangian \fl.

\newsec{Matrix Elements}

The discussion has focused on the applications of reparameterisation
invariance to the effective lagrangian; in this section we discuss some
of the applications to matrix elements in the heavy particle effective
field theory. As might be expected, the only constraint it places on
matrix elements is entirely trivial.  Labelling states with both velocity
and residual momentum increases the number of possible form factors
allowed;  imposing reparameterisation invariance simply reduces these
back to the usual number of form factors.

States in the effective theory have a velocity $v$ and
a residual momentum $k$, with total momentum $p=mv+k$. Thus there is
also a reparameterisation invariance transformation on the physical
states which redefines $v$ and $k$, but keeps $p$ fixed. Consider the
matrix element of the vector current between two spinless particles,
\eqn\vimat{
\bra{v,k'} j^\mu \ket{v,k} = f_1 v^\mu + f_2 \left(k^\mu+ k'^\mu\right)
+ f_3 \left(k^\mu-k'^\mu\right),
}
where $f_i$ are three independent form factors, and
\eqn\vcur{
j^\mu=
\bar\psi_v \left(v^\mu + {i D^\mu\over
m}\right)\psi_v.
}
It is well known that
this matrix element should have only two independent form factors, $f_+$
and $f_-$. The reparameterisation invariance on the states may be used to
show that one can eliminate one of the form factors, and write
Eq.~\vimat\  in the form
\eqn\vimatnew{
\bra{v,k'} j^\mu \ket{v,k} = f_1 \left(v^\mu + {k^\mu+
k'^\mu\over
2m}\right)
+ f_3 \left(k^\mu-k'^\mu\right),
}
where $f_1$ and $f_3$ are functions of $v+k/m$ and $v'+k/m$.
This is equivalent to the $f_{\pm}$ form factor decomposition, and is
a trivial application of reparameterisation invariance; there are
redundant variables in the effective theory which lead to redundant form
factors which can then be eliminated.

Finally, one can easily see that the
formul\ae\ of Secs.~2--3 can be applied to
external states with velocity $v$, residual momentum $k$, and spin, by
replacing $p/m$ by $v + k/m$. There is no operator ordering ambiguity because
the residual momentum $k$ for external states is a number. The redundant form
factors for particles with spin can be eliminated using the methods used above
for the form factors of spinless particles.

\bigskip
\centerline{\bf Acknowledgements}
We would like to thank A.~Falk, E.~Jenkins, M.~Savage and M.~B.~Wise for
useful discussions.
This work was supported in part by DOE grant \doe,
and by a NSF Presidential Young Investigator award \pyiam.
\bigskip

\appendix{A}{Lorentz Boosts}

The Lorentz boost
\eqn\alam{
\Lambda(w,v,\theta) = \exp\left[i J_{\alpha\beta} w^\alpha v^\beta
\theta\right],\qquad
\left[J_{\alpha\beta}\right]_{\mu\nu}=-i\left(g_{\alpha\mu}
g_{\beta\nu}-g_{\alpha\nu} g_{\beta\mu}\right),
}
is a Lorentz boost in the $w-v$ plane with boost parameter $\theta$.
To compute $\Lambda(w,v,\theta)$ explicitly, define the matrix
\eqn\anmat{
N^\alpha{}_\beta = w^\alpha v_\beta - v^\alpha w_\beta,
\qquad\Lambda(w,v,\theta) = e^{\theta N},
}
A straightforward computation by expanding the exponential in a power series
gives
\eqn\alammess{\eqalign{
\Lambda(w,v,\theta)^\alpha{}_\beta =&
g^\alpha{}_\beta + \left({1 - \cosh \lambda\theta\over\lambda^2}\right)
\left(w^\alpha w_\beta + v^\alpha v_\beta\right)
 + {\sinh \lambda\theta\over \lambda}\left( w^\alpha v_\beta
-v^\alpha w_\beta\right)\cr
&+ \left(w \cdot v\right)\left({\cosh\lambda\theta-1\over\lambda^2}\right)
\left(w^\alpha v_\beta + v^\alpha w_\beta\right),
\cr
}}
where
\eqn\aldef{
\lambda^2 = \left(w\cdot v\right)^2 -1.
}
To obtain the boost matrix $\Lambda(w,v)$ which rotates $v$ into $w$, the boost
parameter $\theta$ must have the value
\eqn\athetval{
\sinh\lambda\theta = \lambda,
}
so that
\eqn\alamb{\eqalign{
\Lambda(w,v)^\alpha{}_\beta =&
g^\alpha{}_\beta - {1\over 1 + v \cdot w}
\left(w^\alpha w_\beta + v^\alpha v_\beta\right)+ \left( w^\alpha v_\beta
-v^\alpha w_\beta\right)\cr
& +{v\cdot w\over 1 + v \cdot w}
\left(w^\alpha v_\beta + v^\alpha w_\beta\right).
\cr
}}

The corresponding transformations $\widetilde\Lambda(w,v,\theta)$
and $\widetilde\Lambda(w,v)$ in the spinor representation may be obtained by
using Eq.~\alam, and replacing the Lorentz generators $J^{\alpha\beta}$ by
their values in the spinor representation,
\eqn\aspingen{
J^{\alpha\beta} w_\alpha v_\beta = -{1\over 2} \sigma^{\alpha\beta}
w_\alpha v_\beta = -{i\over 4}\left[\slash w\ , \ \slash v\right].
}
The exponential is evaluated explicitly using the identity
\eqn\aspinexp{
\left[\slash w\ ,\ \slash v\right]^2 = 4 \lambda^2,
}
to give
\eqn\alamspin{
\widetilde\Lambda(w,v,\theta) = \cosh\left({\lambda\theta\over2}\right)
+{1\over 2 \lambda}
\left[\slash w\ , \slash v\right]\sinh\left({\lambda\theta\over2}\right).
}
For the transformation that rotates $v$ into $w$, $\theta$ has the value
Eq.~\athetval, so that
\eqn\alamrot{
\widetilde\Lambda(w,v) =
{1 + \slash w\ \slash v\over \sqrt{2\left(1 + v \cdot w\right)}}.
}

\listrefs
\end